\documentclass[aps,prd,preprint,superscriptaddress]{revtex4-2}
\usepackage{graphicx,bm,float,amsthm,amsmath,xcolor, colortbl,ulem,multirow}
\usepackage[ruled,vlined]{algorithm2e}

\begin{document}
\title{Simulating neutrino oscillations on a superconducting qutrit}
\author{Ha C. Nguyen}
\affiliation{Nano and Energy Center, University of Science, Vietnam National University, Hanoi, Vietnam}
\affiliation{Pheninkaa Institute for Advanced Study, Phenikaa University, Hanoi, 12116, Vietnam}

\author{Bao G. Bach}
\affiliation{Ho Chi Minh City University of Technology, VNU HCM, Vietnam}

\author{Tien D. Nguyen}
\affiliation{Nano and Energy Center, University of Science, Vietnam National University, Hanoi, Vietnam}
\affiliation{Faculty of Physics, Hanoi University of Science, VNU Hanoi, Vietnam}

\author{Duc M. Tran}
\affiliation{Nano and Energy Center, University of Science, Vietnam National University, Hanoi, Vietnam}
\affiliation{Laboratoire ICB, UMR CNRS, Universite de Bourgogne Franche-Comte, Dijon, France}

\author{Duy V. Nguyen}
\affiliation{Pheninkaa Institute for Advanced Study, Phenikaa University, Hanoi, 12116, Vietnam}
\affiliation{Faculty of Computer Science, Phenikaa University, Hanoi, 12116, Vietnam}

\author{Hung Q. Nguyen}
\affiliation{Nano and Energy Center, University of Science, Vietnam National University, Hanoi, Vietnam}
\email{hungngq@hus.edu.vn}

\begin{abstract}
Precise measurements of parameters in the PMNS framework might lead to new physics beyond the Standard Model. However, they are incredibly challenging to determine in neutrino oscillation experiments. Quantum simulations can be a powerful supplementary tool to study these phenomenologies. In today's noisy quantum hardware, encoding neutrinos in a multi-qubit system requires a redundant basis and tricky entangling gates. We encode a three-flavor neutrino in a superconducting qutrit and study its oscillations using PMNS theory with time evolution expressed in terms of single qutrit gates. The qutrit is engineered from the multi-level structure of IBM transmon devices. High-fidelity gate control and readout are fine-tuned using programming microwave pulses using a high-level language. Our quantum simulations on real hardware match well to analytical calculations in three oscillation cases: vacuum, interaction with matter, and CP-violation. 
\end{abstract}
\maketitle

\section{Introduction}
The discovery of neutrino oscillations \cite{SK1998, SNO2002} introduces at least seven parameters to particle physics models, including three masses and four lepton mixing parameters. It also implies lepton mixing, meaning a neutrino flavor is not one but a superposition of mass eigenstates.
Experimental results for these parameters are mainly interpreted in the framework of the 3x3 unitary mixing matrix called the Pontecorvo - Maki - Nakagawa - Sakata (PMNS) theory, in which the three active massive neutrinos $\nu_i$ ($i=1,2,3$) are related to the three known flavors $\nu_\alpha$ ($\alpha = e,\mu,\tau$)  as $|\nu_{\alpha}\rangle=U_{\rm{PMNS}}|\nu_i\rangle$. In the standard three-flavor mixing scheme \cite{Giganti_PPNP2018}, due to the rephasing invariance, the PMNS matrix of Dirac neutrinos is fully described by three mixing angles $\theta_{12}$, $\theta_{23}$, $\theta_{13}$ and a complex phase $\delta$ related the to charge-conjugation and parity-reversal (CP) symmetry violations as
\begin{equation}
    U_{\rm{PMNS}} = 
    \begin{bmatrix} 
        c_{12}c_{13} & s_{12}c_{13} & s_{13}e^{-i\delta} \\
        -s_{12}c_{23} - c_{12}s_{23}s_{13}e^{i\delta} & c_{12}c_{23} - s_{12}s_{23}s_{13}e^{i\delta} & s_{23}c_{13} \\
        s_{12}s_{23} - c_{12}c_{23}s_{13}e^{i\delta} & -c_{12}s_{23} - s_{12}c_{23}s_{13}e^{i\delta} & c_{23}c_{13}
    \end{bmatrix}, \label{eq:UPMNS0}
\end{equation}
with $c_{ij}=\cos\theta_{ij}$ and $s_{ij}=\sin\theta_{ij}$. In essence, neutrino flavors' spontaneous transformation is a quantum interference phenomenon due to the wave nature of neutrinos with their mass eigenstates time-dependently acquiring different phases. The dynamic of neutrino oscillations is governed by a unitary Hamiltonian, which separates into the kinetic and potential parts $H=H_0+H_1$ \cite{DentonPhysRevD19} as
\begin{equation}
    H_0 = \frac{1}{2E}U_{\rm{PMNS}} 
    \begin{bmatrix} 
        0 & 0 & 0  \\
        0 & \Delta m_{21}^2 & 0 \\
        0 & 0 & \Delta m_{31}^2 \\
    \end{bmatrix} U_{\rm{PMNS}}^{\dagger},
\end{equation}
and
\begin{equation}
    H_1 = \frac{1}{2E}
    \begin{bmatrix} 
        V_m & 0 & 0 \\
        0 & 0 & 0 \\
        0 & 0 & 0 \\
    \end{bmatrix}.\label{Matrix:Vc}
\end{equation}
Here $\Delta m^2_{ij}=m_{i}^2-m_{j}^2$ are the neutrino mass-squared differences, and $V_m$ is the Wolfenstein matter potential \cite{WolfensteinPRD}. This potential stems from the coherent forward elastic scattering with the matter electrons and is written in unit eV$^2$.

The matter interaction can be considered a perturbation problem. To maintain a similar form compared to the vacuum case, the Hamiltonian is diagonalized \cite{DentonJHEP2016, neutrino2019glass} as
\begin{equation}\label{PMNS:matter}
     H = \frac{1}{2E} U_{\rm{PMNS}}(\hat\theta_{12},\hat\theta_{23},\hat\theta_{13}) 
    \begin{bmatrix} 
        0 & 0 & 0  \\
        0 & \Delta \hat m_{21}^2 & 0 \\
        0 & 0 & \Delta \hat m_{31}^2 \\
    \end{bmatrix} U_{\rm{PMNS}}^{\dagger}(\hat\theta_{12},\hat\theta_{23},\hat\theta_{13}),
\end{equation}
with the hat denoting parameters related to the matter interaction case. Here,
\begin{equation}
    \Delta \hat m^2_{21} = \Delta m^2_{21}\sqrt{(\cos2\theta_{12}-a_{12}/\Delta m^2_{21})^2 + \cos^2(\theta_{13}-\hat\theta_{13})\sin^22\theta_{12}},
\end{equation}
and
\begin{equation}
    \Delta \hat m^2_{31} = \Delta m^2_{31}+\frac14 V_m+\frac12 (\Delta \hat m^2_{21}-\Delta m^2_{21})+\frac34 (\Delta \hat m^2_{ee}-\Delta m^2_{ee})
\end{equation}
are associated with the energy levels in matters. $a_{12} = \frac12(V_m+\Delta m^2_{ee}-\Delta \hat m^2_{ee})$, where $\Delta m^2_{ee} = c_{12}^2\Delta m^2_{31}+s_{12}^2\Delta m^2_{32}$ is the effective mass-squared difference, and its corresponding quantity in matters is $\Delta \hat m^2_{ee} = \Delta m^2_{ee}\sqrt{(\cos2\theta_{13}-V_m/\Delta m^2_{ee})^2+\sin^22\theta_{13}}$ \cite{StephenPRD2016}. The original $\theta_{ij}$ is modified into \cite{DentonJHEP2016}
\begin{align}
    \sin\hat\theta_{12} &= \sqrt{\frac{1}{2}-(\Delta m_{21}^2\cos2\theta_{12}-a_{12})/2\Delta \hat m_{21}^2},\label{thetahat1}\\
    \sin\hat\theta_{13} &= \sqrt{\frac{1}{2}-(\Delta m^2_{ee}\cos2\theta_{13}-V_m)/2\Delta \hat m^2_{ee}},\label{thetahat2}\\
    \hat\theta_{23} &= \theta_{23}.\label{thetahat3}
\end{align}

Similar to the case of vacuum or CP-violation oscillations, the Hamiltonian in these scenarios has a diagonal form. From an arbitrary initial state $|\nu(0)\rangle \equiv |\nu_\alpha\rangle = \sum_i U^*_{\alpha i}|\nu_i\rangle,$ the neutrino evolves in time $t$ in the matrix form as
\begin{align}\label{UPMNS}
    |\nu(t)\rangle &= e^{-iHt} |\nu(0)\rangle\notag\\
    &= U_{\rm{PMNS}} \Lambda(t) U^{\dagger}_{\rm{PMNS}}|\nu(0)\rangle\notag \\
    &=U_{\rm{PMNS}} \begin{bmatrix} 
        1 & 0 & 0  \\
        0 & e^{-i\Delta m_{21}^2\frac{t}{2E}} & 0 \\
        0 & 0 & e^{-i\Delta m_{31}^2\frac{t}{2E}} \\
    \end{bmatrix} U^{\dagger}_{\rm{PMNS}}|\nu(0)\rangle,
\end{align}
with $E$ as the neutrino energy. Eq.~\eqref{UPMNS} is equivalent to 
\begin{equation}
|\nu(t)\rangle
= \sum_i U^*_{\alpha i}e^{-i m_i^2\frac{t}{2E}}|\nu_i\rangle =
\sum_i U^*_{\alpha i}e^{-i m_i^2\frac{t}{2E}}\sum_\beta U_{\beta i}|\nu_\beta\rangle.
\end{equation}
Here, $U^*_{\alpha i}$ denotes the corresponding terms in the PMNS matrix. The probability of detecting neutrino oscillations from flavor $\alpha$ to $\beta$ is
\begin{equation}\label{eq:classical_prob}
    P_{\alpha\rightarrow\beta}=|\langle \nu_\beta|\nu(t)\rangle|^2=\big|\sum_i U^*_{\alpha i} U_{\beta i} e^{-i m^2_i\frac{t}{2E}}\big|^2.
\end{equation}

The PMNS theory has been verified experimentally via different sources of neutrino fluxes \cite{snowmass2022}. The parameters that are reasonably well measured are the solar mixing angle $\theta_{12} \approx 34^\circ$ \cite{KamLAND}, the reactor mixing angle $\theta_{13} \approx 8.5^\circ$ \cite{DayaBay, RENO, DoubleChooz}, and the solar mass splitting $\Delta m_{21}^2 \approx 7.5 \times10^{-5}\text{eV}^2$ \cite{KamLAND}. The two parameters with well-determined partial information are the atmospheric mixing angle $\theta_{23} \approx 45^\circ$ and the atmospheric mass splitting $\Delta m_{31}^2 \approx \pm 2.5 \times 10^{-3}\text{eV}^2$ \cite{snowmass2022}. However, the phase $\delta$ in CP-violation still needs to be discovered with significant certainties \cite{T2KNat2020}. Excellent control and more data from further accelerator experiments such as HK \cite{HyperK} and DUNE \cite{Dune2016} are required to suppress systematic experimental errors. Additional constraints on the value of the complex phase would establish or deny the CP violation in the lepton sector that might explain the matter–antimatter disparity through leptogenesis \cite{T2KNat2020}. The matter effects from natural or artificial sources are crucial when the interactions between neutrinos and electrons, protons, and neutrons are large. This is the case in accelerator experiments where matter effects of the earth give rise to spurious CP asymmetry \cite{CPfakematter}.

On the other hand, the PMNS framework can be studied by tools from quantum simulation. Recent progress in quantum engineering has realized noisy intermediate-scale quantum (NISQ) computers, devices that perform key proof-of-concept quantum algorithms and showcase enormous potential \cite{GoogleSupremacy, ChinaSupremacy}. Designed as a universal computing platform and programmed using high-level language through cloud access, there is a range of remarkable works performed on real devices using a small number of qubits. They include demonstrating critical quantum algorithms \cite{ShorIBM, GroverIBM}, simulation of quantum phenomena \cite{SolanoIon, MartinisNComm, BlattNature, HouckQED, Gambetta17, Tran2022}, or reproducing foundation quantum experiences \cite{IBMBell, IBMduality, IBMcat, DevittPRA}.

In certain problems, multi-level structures of qudits utilizing larger computational spaces are promising architectures for quantum computations and simulations \cite{GustafsonPRD2021, CiavarellaPRD2021}. While the higher energy level is more prone to noise, qutrits have been realized successfully on various hardware architectures, especially superconducting circuits \cite{BlaisRevModPhys21, KrantzAPR2019, KwonAIP21, RasmussenPRXQuantum21, GaoPRXQuantum21}. A range of physics phenomena have been simulated on its most popular platform, the transmon \cite{transmon}, including efficient quantum gates \cite{ChuNatPhy2022}, quantum information scrambling \cite{BlokPRX21}, topological phase transition \cite{TanPhysRevLett18}, sensors for microwave fields \cite{KristenNPJ2020}, quantum number generators \cite{Kulikov_2017}, contextuality without nonlocality \cite{JergerNatCom2016}, or quantum metrology \cite{NikolaevaPRA22}.

The quantum coherence of oscillating neutrinos over long distances provides a natural system for quantum simulations. The PMNS matrix has been solved for multi-qubit systems \cite{ArguellesPRR19, MolewskiPRD22, Jha2022} that set the basis to simulate neutrino time evolution. Initially demonstrated on superconducting hardware \cite{ArguellesPRR19}, a PMNS matrix is parameterized for the case of vacuum oscillations using two qubits. The PMNS qubit parameterization is then improved to include CP-violation \cite{MolewskiPRD22}, but its circuit is too complicated for current NISQ hardware. Furthermore, encoding three neutrino flavors on a 2-qubit system requires a redundant basis state and suffers from large errors in entangling gates. This mismatch is unavoidable in many high-energy physics problems where the local degrees of freedom are not even number \cite{HEP2022}. The inefficiency of qubit mapping hinders current attempts to simulate more complex phenomena. Indeed, quantum simulations are only presented in the two-flavor picture of collective neutrino oscillations in core-collapse supernovae \cite{HallPRD21, YeterAydenizQIP2022}. These issues necessitate a more NISQ-efficient encoding for high-energy physics simulations using three-level systems, such as qutrits.

In this work, we simulate neutrino oscillations by harnessing the computational power of high dimensional Hilbert space while maintaining a low circuit depth. The three neutrino flavors are encoded in a transmon qutrit, and their quantum oscillations are simulated following the PMNS theory. The original PMNS matrix is decomposed into native qutrit gates in three cases: bare vacuum oscillations, oscillations with matter interaction $V_m \neq 0$, and oscillations with CP-violation $\delta \neq 0$. Using the Pulse package in IBM's Qiskit, low-level microwave pulses are engineered to access the qubit's third level, thus constructing a transmon qutrit. Distinctly discriminated between their levels, the qutrit has high-quality gates. By carefully tracking phase advances between the two subspaces \{01\} and \{12\}, the oscillations simulated on the transmon qutrit match well to analytical results obtained from PMNS theory in all three scenarios. In all calculations, values from experiments are used according to the normal mass hierarchy in NuFIT 5.1 data \cite{NUFIT}: $\theta_{12}=33.45^\circ$, $\theta_{23}=42.1^\circ$, $\theta_{13}=8.62^\circ$, $\Delta m_{21}^2=7.42\times10^{-5}~\text{eV}^2$, and $\Delta m_{31}^2 = 2.510\times10^{-3}~\text{eV}^2$.
 
\section{Method}
\subsection{The PMNS theory as qutrit gate decompositions}
\begin{figure}[ht]
    \begin{center}
    \includegraphics[width=0.8 \textwidth]{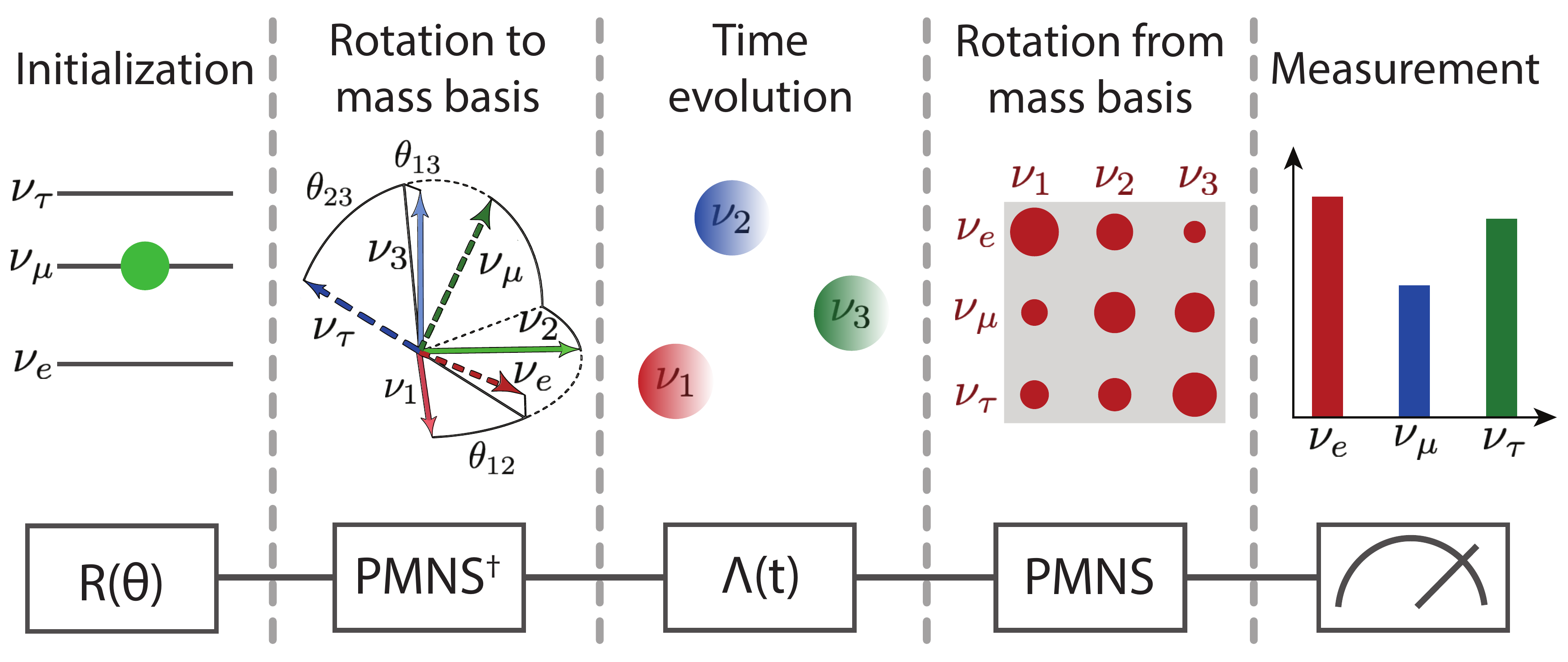}
    \caption{\textbf{The PMNS theory for neutrino oscillations:} with non-degenerate masses and flavor mixings, a neutrino unitarily transforms between its mass and flavor bases. This PMNS transformation is graphically represented by Euler angles between the two bases, or equivalently, a non-diagonal matrix. The measurement outcome depends on time evolution, which is proportional to the traveling distances. The second row sketches this process using quantum circuit language. First, a unitary qutrit gate $\mathcal R^3(\theta)$ as in Eq. ~\eqref{gate:general} initiates the quantum state from $|0\rangle$, then a gate combination  $R^{\dagger}_{\rm{PMNS}}$ as in Eq.~\eqref{RPMNS2} rotates it to the mass basis. This state $|\nu(t)\rangle$ evolves in time, equivalent to the action of phase gates in the two subspaces. Before measurement, $|\nu(t)\rangle$ is rotated back to the flavor basis using $R_{\rm{PMNS}}$ as in Eq.~\eqref{RPMNS1}.}\label{fig:neutrino}
    \end{center}
\end{figure}

The three neutrino flavor eigenstates are directly mapped to orthogonal states of a three-level qutrit as
\begin{equation}
    |\nu_{e}\rangle = |0\rangle \equiv
    \begin{pmatrix}
        1 \\ 0 \\ 0
    \end{pmatrix},\hspace{10pt}
    |\nu_{\mu}\rangle = |1\rangle \equiv
    \begin{pmatrix}
        0 \\ 1 \\ 0
    \end{pmatrix},\hspace{10pt}
    |\nu_{\tau}\rangle =  |2\rangle \equiv
    \begin{pmatrix}
        0 \\ 0 \\ 1
    \end{pmatrix}.
\end{equation}
Hence, an arbitrary neutrino state is written as a superposition of these eigenstates $|\nu_\alpha\rangle=c_0|0\rangle+c_1|1\rangle+c_2|2\rangle$, where $c_0, c_1, c_2$ are complex numbers satisfying the normalization condition $|c_0|^2+|c_1|^2+|c_2|^2=1$. The transformation to and from the mass basis is done by a rotation under the PMNS-like action $|\nu_\alpha\rangle = U_{\rm{PMNS}}|\nu_i\rangle$, which can be decomposed into rotations within subspaces \{01\}, \{02\} and \{12\}.

To execute the PMNS action on a qutrit, it needs to be decomposed into pulsable gates that can implement with the Pulse package in Qiskit. The PMNS matrix Eq.~\eqref{eq:UPMNS0} can be rewritten in the form
\begin{equation}
    U_{\rm{PMNS}} = 
    \begin{bmatrix} 
        1 & 0 & 0 \\
        0 & c_{23} & s_{23}\\
        0 & -s_{23} & c_{23} \\
    \end{bmatrix}
    \begin{bmatrix} 
        c_{13} & 0 & s_{13}e^{-i\delta} \\
        0 & 1 & 0\\
        -s_{13}e^{i\delta} & 0 & c_{13} \\
    \end{bmatrix}
    \begin{bmatrix} 
        c_{12} & s_{12} & 0 \\
        -s_{12} & c_{12} & 0\\
        0 & 0 & 1 \\
    \end{bmatrix},\label{eq:UPMNS3}
\end{equation}
where the last matrix represents a rotation with $\theta_{12}$. In our qutrit representation, this last term is equivalent to a rotation of an angle $2\theta_{12}$ in subspace \{01\}.

In general, an arbitrary unitary 3x3 qutrit gate $\mathcal R^3$ can be decomposed into rotations in their \{01\} or \{12\} subspace \cite{KononenkoPRR21} as
\begin{equation}\label{gate:general}
    \mathcal R^3=X_0 R^{01}_{\phi_1}(\theta_1) R^{12}_{\phi_2}(\theta_2) R^{01}_{\phi_3}(\theta_3),
\end{equation}
where $X_0$ is diagonal in the computational basis and
\begin{equation}\label{GivenRmn}
    R^{mn}_\phi(\theta)=\exp[-i\frac{\theta}{2}(\sigma_x^{mn}\cos\phi +\sigma_y^{mn}\sin\phi)].
\end{equation}
The superscripts denote subspaces of the gate, $\sigma_x^{mn}=|m\rangle\langle n| + |n\rangle\langle m|$, $\sigma_y^{mn}=i(|n\rangle\langle m| -|m\rangle\langle n|)$, $\theta$ is the angle and $\phi$ is the axis of the rotation. Clearly, there is no unique decomposition of a general qutrit gate. In practice, we decompose only the first two matrices in Eq.~\eqref{eq:UPMNS3} and require that it takes the form $R^{01}R^{12}R^{01}$. This way, the number of gates is minimal to avoid systematic errors on NISQ hardware.

In the simplest scenario when there is no CP violation $\delta=0$ and no matter interaction $V_m=0$, the original PMNS matrix $U_{\rm{PMNS}}$ in Eq.~\eqref{eq:UPMNS3} is decomposed as combinations of qutrit gates in their \{01\} and \{12\} subspaces as
\begin{equation}\label{RPMNS1}
    R_{\rm{PMNS}} = R^{01}_{\frac{\pi}{2}}(\alpha_1)R^{12}_{\frac{3\pi}{2}}(\alpha_2)R^{01}_{\frac{\pi}{2}}(\alpha_3)R^{01}_{\frac{\pi}{2}}(-2\theta_{12}). 
\end{equation}
Its conjugate writes
\begin{equation}\label{RPMNS2}
    R^{\dagger}_{\rm{PMNS}} = R^{01}_{\frac{\pi}{2}}(2\theta_{12})R^{01}_{\frac{\pi}{2}}(-\alpha_3)R^{12}_{\frac{3\pi}{2}}(-\alpha_2)R^{01}_{\frac{\pi}{2}}(-\alpha_1). 
\end{equation}
Here, $\alpha_i$ relates to $\theta_{ij}$ as 
\begin{align}
    \cos\frac{\alpha_1}{2} & = - \frac{\cos\theta_{13}\sin\theta_{23}}{\sqrt{1-\cos^2\theta_{13}\cos^2\theta_{23}}},\label{eq:alpha1}\\
    \cos\frac{\alpha_2}{2} & = \cos\theta_{13}\cos\theta_{23},\label{eq:alpha2}\\
    \cos\frac{\alpha_3}{2} & = - \frac{\sin\theta_{23}}{\sqrt{1-\cos^2\theta_{13}\cos^2\theta_{23}}}.\label{eq:alpha3}
\end{align}

In our quantum circuits, the time operator $\Lambda(t)$ is constructed by modifying the phases of the following pulses by the argument difference between their two entries \cite{KononenkoPRR21}. In particular, after the diagonal gate, a phase of 
\begin{equation}\label{Phi01}
    \Phi^{01}=-\Delta m_{21}^{2}\frac{t}{2E} 
\end{equation}
is added to gates in subspace \{01\}, and a phase of 
\begin{equation}\label{Phi12}
    \Phi^{12}
    =\Delta m_{21}^{2}\frac{t}{2E} -\Delta m_{31}^{2}\frac{t}{2E}
    =-\Delta m_{32}^{2}\frac{t}{2E} 
\end{equation}
is added to gates in subspace \{12\}. In the relativistic scale, $t = L$ the traveling distance. The time operator $\Lambda(t)$ can be written as $\Lambda(L/E)$, which is more prevalent in the neutrino community. These two phases are linearly related with constant neutrino mass-squared differences measured from experiments. At the end of the circuit, we perform a change back to the flavor eigenstates basic and measure the probability of neutrino flavors. All in all, the combined gate that drives the qutrit is
\begin{align}\label{RPMNS:vac}
    \mathcal R^3_0 &= R_{\rm{PMNS}} \Lambda(L/E) R^{\dagger}_{\rm{PMNS}}\notag\\
    &=R^{01}_{\frac{\pi}{2}+\Phi^{01}}(\alpha_1)R^{12}_{\frac{3\pi}{2}+\Phi^{12}}(\alpha_2)R^{01}_{\frac{\pi}{2}+\Phi^{01}}(\alpha_3-2\theta_{12})R^{01}_{\frac{\pi}{2}}(-\alpha_3+2\theta_{12})R^{12}_{\frac{3\pi}{2}}(-\alpha_2)R^{01}_{\frac{\pi}{2}}(-\alpha_1).
\end{align}
Measuring this state yields the probability distribution of an oscillating neutrino in a vacuum, as the PMNS mechanism dictates. 

A similar decomposition is applied with the diagonalized PMNS Hamiltonian Eq.~\eqref{PMNS:matter} for the case of oscillations with matter interaction. Without CP-violation $\delta=0$, the decomposition writes
\begin{equation}
    R_{\rm{PMNS}}(\hat\theta_{12},\hat\theta_{23},\hat\theta_{13}) = R^{01}_{\frac{\pi}{2}}(\hat{\alpha}_1)R^{12}_{\frac{3\pi}{2}}(\hat{\alpha}_2)R^{01}_{\frac{\pi}{2}}(\hat{\alpha}_3-2\hat{\theta}_{12}). 
\end{equation}
Here $\hat\alpha_i$ depends on $\hat\theta_{ij}$ in a similar manner to $\alpha_i$ depends on $\theta_{ij}$ following Eq.~\eqref{eq:alpha1}, \eqref{eq:alpha2}, and \eqref{eq:alpha3}. The matter-related term $\hat\theta_{ij}$ relates to $\theta_{ij}$ according to Eq.~\eqref{thetahat1}, \eqref{thetahat2}, and \eqref{thetahat3}. Using the same decomposition as in Eq.~\eqref{RPMNS1} and \eqref{RPMNS2}, the quantum circuit is identical to the vacuum case with matter equivalents replacing vacuum mixing parameters 
\begin{equation}\label{R:matter}
    \hat{\mathcal R^3} = R^{01}_{\frac{\pi}{2}+\hat\Phi^{01}}(\hat\alpha_1)R^{12}_{\frac{3\pi}{2}+\hat\Phi^{12}}(\hat\alpha_2)R^{01}_{\frac{\pi}{2}+\hat\Phi^{01}}(\hat\alpha_3-2\hat\theta_{12})R^{01}_{\frac{\pi}{2}}(-\hat\alpha_3+2\hat\theta_{12})R^{12}_{\frac{3\pi}{2}}(-\hat\alpha_2)R^{01}_{\frac{\pi}{2}}(-\hat\alpha_1).
\end{equation}
It is straightforward to generalize this formula to the case of oscillations with matter interaction with CP-violation $\delta\neq 0$ and other similar scenarios with a diagonalized Hamiltonian.

In the present of CP broken symmetry $\delta\neq 0$, the PMNS matrix $U_{\rm{PMNS}}$ contains a complex phase $\delta$ associated with the $\sin(\theta_{13})$ term. Following the strategy used to derive Eq.~\eqref{RPMNS1} and \eqref{RPMNS2}, the decomposition of the PMNS and its conjugate in the case of nonzero $\delta$ are
\begin{align}\label{RPMNS:delta}
R_{\rm{PMNS}} &= R^{01}_{\frac{\pi}{2}+\delta}(\alpha_1)R^{12}_{\frac{3\pi}{2}}(\alpha_2)R^{01}_{\frac{\pi}{2}+\delta}(\alpha_3)R^{01}_{\frac{\pi}{2}}(-2\theta_{12}),\\
R_{\rm{PMNS}}^{\dagger} &= R^{01}_{\frac{\pi}{2}}(2\theta_{12})R^{01}_{\frac{\pi}{2}+\delta}(-\alpha_3)R^{12}_{\frac{3\pi}{2}}(-\alpha_2)R^{01}_{\frac{\pi}{2}+\delta}(-\alpha_1).
\end{align}
Here $\alpha_1,\alpha_2,\alpha_3$ are determined from Eq.~\eqref{eq:alpha1},\eqref{eq:alpha2}, and \eqref{eq:alpha3}. We note that this decomposition contains four qubit gates and is not minimal. It can be further reduced to 3 rotations by combining the two adjacent gates in \{01\} subspace. Nevertheless, it reduces to Eq.~\eqref{RPMNS1} and \eqref{RPMNS2} when $\delta = 0$. The evolution of the qutrit follows
\begin{align}\label{R:delta}
    \mathcal R^3_{\delta} &=R^{01}_{\frac{\pi}{2}+\delta+\Phi^{01}}(\alpha_1)R^{12}_{\frac{3\pi}{2}+\Phi^{12}}(\alpha_2)R^{01}_{\frac{\pi}{2}+\delta+\Phi^{01}}(\alpha_3)R^{01}_{\frac{\pi}{2}+\Phi^{01}}(-2\theta_{12})\notag\\
    &\times R^{01}_{\frac{\pi}{2}}(2\theta_{12})R^{01}_{\frac{\pi}{2}+\delta}(-\alpha_3)R^{12}_{\frac{3\pi}{2}}(-\alpha_2)R^{01}_{\frac{\pi}{2}+\delta}(-\alpha_1),
\end{align}
with $\Phi^{ij}$ is defined in Eq.~\eqref{Phi01} and \eqref{Phi12}. This action simulates the evolution of neutrinos in the presence of CP violation $\delta\neq 0$.

\subsection{Control and readout of the transmon qutrit}

\begin{figure}[ht]
    \begin{center}
    \includegraphics[width=0.8\textwidth]{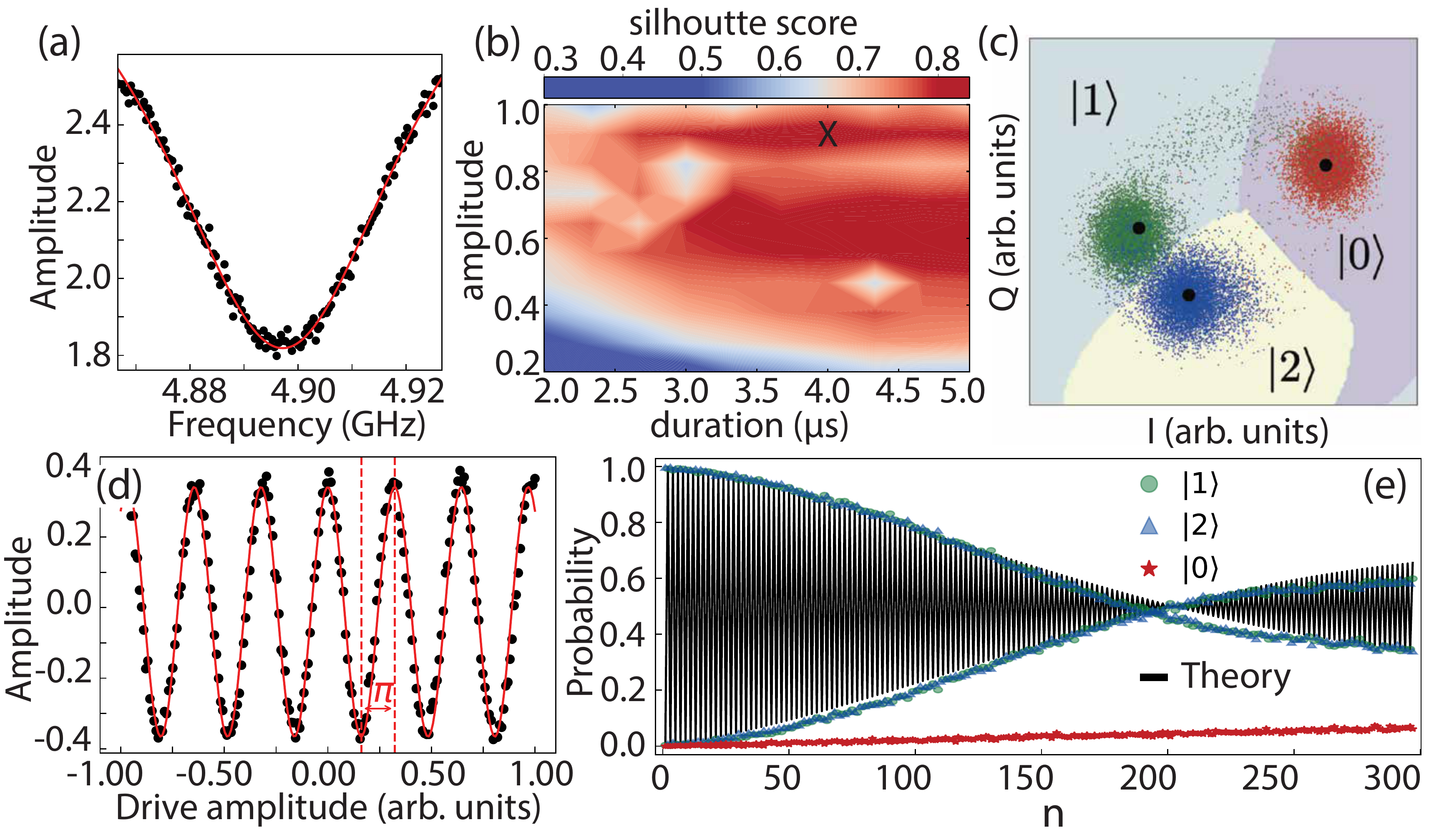}
    \caption{\textbf{Engineering the qutrit:}(a) Spectroscopy results of qubit 0 on IBM's Jakarta device with the resonance peak between state $|1\rangle\leftrightarrow|2\rangle$ $f^{12}=4.897$ GHz. (b) The silhouette score heat map as a function of durations and amplitudes of the measurement pulses. The marker highlights the qutrit readout sweet spot with the highest score. (c) Qutrit discriminator trained by SVC algorithm from data of state preparation of the three lowest energy levels. (d) Rabi oscillations in subspace \{12\}. Amplitude for $\pi$ pulse $R_X^{12}(\pi)$ is defined as half of the period marked by the two dash lines. (e) Error amplifying of a $R_X^{12}(\pi)$ pulse using protocol $[R_X^{12}(\pi)]^nR_X^{01}(\pi)$. The $R_X^{12}(\pi)$ pulse is applied $n$ times on a prepared state $|1\rangle$ \cite{gateerrors}. The black line shows the fit function to the probability of state $|1\rangle$ that indicates calibration errors.}\label{fig:qutrit}
    \end{center}
\end{figure}

\begin{table}[ht]
\begin{tabular}{||c|c|c||}\hline
Properties  & Symbol & Value \\\hline
Qutrit frequency $|0\rangle \leftrightarrow |1\rangle$ & $f^{01}$ & 5.237 GHz \\\hline
Qutrit frequency $|1\rangle \leftrightarrow |2\rangle$ & $f^{12}$ & 4.897 GHz \\\hline
Pulse resolution & $dt$ & 0.222 ns  \\\hline
Lifetime $(|0\rangle \leftrightarrow |1\rangle$) & $T_1$ &  184.5 $\mu$s \\\hline
Coherence time ($|0\rangle \leftrightarrow |1\rangle$) & $T_2$ & 40.39 $\mu$s   \\\hline
Readout pulse length &  & 4 $\mu$s \\\hline
Transmon regime & $E_J/E_C$ & 33.65 \\\hline
\end{tabular}
\caption{\textbf{Properties of qubit 0 on IBM Jakarta device} obtained in March 2022. This seven qubit device has a standard transmon architecture with I shape connectivity.}\label{tab:IBM}
\end{table}

To implement the above decompositions on real hardware, we modify IBM's transmon qubit into a qutrit using Qiskit's Pulse package. By changing the amplitude, phase, and duration of a time-dependent pulse using high-level scripts written in Python, an arbitrary waveform can be constructed to access and control the transmon's third level. We focus on qubit 0 of the Jakarta device, a seven-qubit machine with detailed information provided in table \ref{tab:IBM}. Further tests on other devices, such as Armonk, yield similar results. This machine has higher systematic errors compared to Jakarta and its results are only reported in our GitHub repository \cite{Github}. In most cases, IBM's default values for \{01\} subspace are used, and we focus on engineering pulses in the \{12\} subspace.

\normalem
\begin{algorithm}[ht]\label{Rabi}
    \SetAlgoLined
    \KwResult{Resonance frequency to the third state and $R_X^{12}(\pi)$ gate}
    Step 1: Transition from $|0 \rangle$ to $|1\rangle$:\\
        \texttt{set\_frequency = $f^{01}$\\
        play Gaussian(dur = $T_d$, ampl = $A^{01}_{\pi}$ )}\\
    Step 2: Find \{12\} transitional frequency $f^{12}$\\
    \texttt{ \For{$f^{12}$ \textbf{in} frequency\_guess}{
        set\_frequency = $f_{12}$\\
        play Gaussian(dur = $T_d$, ampl = ampl12 ) } }
    Lorentzian fit $\rightarrow$ resonance peak $\rightarrow$ $f^{12}$\\
    Step 3: Perform Rabi experiment in subspace \{12\} to find $R_X^{12}(\pi)$  \\
    \texttt{ \For {amp12 \textbf{in} amplitudes\_12} {
        set\_frequency = $f^{12}$\\
        play Gaussian(dur = $T_d$, ampl = ampl12)} } 
        % set\_frequency = $f^{01}$: play $R_X^{01}(\pi)$\\
        % set\_frequency = $f^{12}$: play Gaussian(dur = $T_d$, ampl = ampl12)} }
    Cosine function fit $\rightarrow$ $A_\pi^{12}$ 
\caption{Accessing the third level of the qutrit}
\end{algorithm}

To define the qutrit, an algorithm that follows the Rabi spectroscopy protocol is implemented, as shown in Fig.~\ref{fig:qutrit} a, d and algorithm \ref{Rabi}. While using IBM's default frequency $f^{01}$ for subspace \{01\}, we find $f^{12}$ by sweeping the frequency anharmonicity to spot the peak of excitation from state $|1\rangle$ to $|2\rangle$. The qubit is first set to the $|1\rangle$ state using a $\pi$ pulse $R^{01}_X(\pi)$. The subspace \{12\} is then allocated by searching for the resonant frequency $f^{12}$ using \texttt{set\_frequency} in Pulse and a Lorentzian fit. A sinusoidal side-band at anharmonicity $f^{12}-f^{01}$ is applied to amplitude-modulated microwave pulses to effectively implement transitions $|1\rangle \leftrightarrow |2\rangle$ \cite{AlbaPRA22}. This Rabi experiment will then return the amplitude of the $\pi$ pulse in subspace \{12\}. 

\begin{algorithm}[ht]\label{Readout}
    \SetAlgoLined
    \KwResult{Improved qutrit discriminator using support vector classification}
    \texttt{\For{ampl \textbf{in} amplitudes}{
    \For{dur \textbf{in} durations}
        {\For{shot \textbf{in} shots}{
        Initialize $|N\rangle$ /* $N = 0, 1, 2$ */\\}
        play measurement\textunderscore pulse([dur,ampl]) \\
        distance = silhouette\textunderscore score( $(|0\rangle, |1\rangle, |2\rangle)$, [dur,ampl] )\\
}}
max(distance) $\rightarrow$ [dur,ampl] }
\caption{Measurement and readout optimization}
\end{algorithm}

To classify output from IBM, we build qutrit discriminators from three-state preparation experiments, in which the durations and the amplitudes of the measurement pulses are tuned. In this experiment, the transmon is repeatedly initialized at states $|0\rangle, |1\rangle$, or $|2\rangle$ and then measured, as outlined in algorithm \ref{Readout}. The output signals rendered from Qiskit level-1 kernelled data are complex numbers $I+iQ$ in the in-phase - quadrature plane. As shown in Fig. 2c, the discriminator is a graph of three clusters corresponding to these prepared states with quality depending on the distance between separate clusters. We use the silhouette-score metric from the Scikit-learn library in Python to quantify these inter-distances. A heat map of the silhouette score as a function of pulse amplitudes and durations is generated to find the sweet spot for the qutrit performance. In Fig.~\ref{fig:qutrit} b, the amplitude is in the range of 0.4-1 in a normalized unit, and the duration sweeps from 2 to 5 $\mu$s. The optimal measurement spot is defined in association with an [amplitude, duration] pair that produces the highest score. We find an optimal measurement pulse with duration 4 $\mu$s and amplitude 0.91 in the normalized unit at the marked location on Fig.~\ref{fig:qutrit} b. From these data, a support vector classification (SVC) is applied to train and define the boundaries of these three regions \cite{SVM}. Subsequently, data are classified as $|0\rangle, |1\rangle$, or $|2\rangle$ based on their location in the I-Q plane. Associated with the discriminator shown in Fig.~\ref{fig:qutrit} c, a typical readout accuracy for state preparation for states $|0\rangle$, $|1\rangle$, and $|2\rangle$ are 98.5\%, 94.3\%, and 94.5\%, respectively. To minimize the state preparation and measurement (SPAM) error, these probabilities are further adjusted using the inverse confusion matrix following the error mitigation protocol \cite{Gambetta17}. To tackle the instability and drifting of the transmons, we build a specific discriminator associated with each job. Each run is corrected with a designated mitigation matrix obtained before any operations.

\begin{algorithm}[ht]\label{Gate}
\SetAlgoLined
\KwResult{An arbitrary single qutrit gate using Eq.~\eqref{gate:general}}
\SetKwFunction{FRot}{$R^{01}$}
\SetKwProg{Fn}{Def}{:}{}
  \Fn{\FRot{$\phi$, $\theta$}}{
  \texttt{set\_frequency = $f^{01}$\\
    phase\_offset = $\phi$ \\
    play Gaussian(dur=$T_d$, ampl=$\frac{\theta}{\pi}A_\pi^{01}$)}\\
    \KwRet $R^{01}$}
\SetKwFunction{FRot}{$R^{12}$}
\SetKwProg{Fn}{Def}{:}{}
    \Fn{\FRot{$\phi$, $\theta$}}{
    \texttt{set\_frequency = $f^{12}$\\
    phase\_offset = $\phi$ \\
    play Gaussian(dur=$T_d$, ampl=$\frac{\theta}{\pi}A_\pi^{12}$)}\\
    \KwRet $R^{12}$ }

    $\mathcal R^3=X_0 R^{01}_{\phi_1}(\theta_1) R^{12}_{\phi_2}(\theta_2) R^{01}_{\phi_3}(\theta_3)$
\caption{Single qutrit gates}
\end{algorithm}

The building block for a universal single qutrit gate is the Given rotation $R^{mn}_\phi(\theta)$ in subspace $\{mn\}$ as defined in Eq.~\eqref{GivenRmn} whose angle $\theta$, and axis of rotation $\phi$ are generated by a pulse with corresponding values of phase and envelope area at resonant frequency $f^{mn}.$ Since the default rotation gates in IBM's Qiskit are built from two square root gates, it is hard to track phases with this protocol. We rebuild rotation gates in both \{01\} and \{12\} with matrix form
\begin{align}
    R_\phi^{01}(\theta) & =
    \begin{bmatrix}
        \cos\theta/2 & -i\sin\theta/2e^{-i\phi} &0\\
        -i\sin\theta/2e^{i\phi} & \cos \theta/2 &0\\
        0 & 0 & 1 
    \end{bmatrix} = 
    \begin{bmatrix} 
        0 & -i &0\\
        -i & 0 &0\\
        0 & 0 & 1 
    \end{bmatrix},\\
    R_\phi^{12}(\theta) & =
    \begin{bmatrix} 
        1 & 0 & 0 \\
        0 & \cos \theta/2 & -i\sin\theta/2e^{-i\phi}\\
        0 & -i\sin\theta/2e^{i\phi} & \cos \theta/2 \\
    \end{bmatrix} = 
    \begin{bmatrix} 
        1 & 0 & 0 \\
        0 & 0 & -i\\
        0 & -i & 0\\
    \end{bmatrix}.
\end{align}
Here, the second equal signs denote the $\theta = \pi$ rotation. To physically execute this $\pi$ rotation, a Gaussian pulse of the form
\begin{equation}
    \Omega(t)=\Omega_0\exp\left[-\frac{(t-T_d/2)^2}{2\sigma^2}\right],
\end{equation}
with mean duration $T_d = 160~dt = 35.56$ ns and deviation $\sigma = 40~dt = 8.89$ ns. The pulse amplitude  $\Omega_0$ is varied by small increments with fixed duration at the resonance frequency of the corresponding subspace. The obtained Rabi oscillation has a sinusoidal form in which its amplitude represents the fraction of the shots driving the qutrit between the two states. The amplitude for $\pi$ pulse $A_\pi$ equals half of the period, as marked by two dash lines in Fig.~\ref{fig:qutrit} d. The angle of an arbitrary rotation $\theta$ is obtained by linearly scaling the envelope area of the $\pi$-pulse by its amplitude $A(\theta)=\frac{\theta}{\pi}A(\pi)$ \cite{smith2022programming}. To modify the rotation axis, we adjust the phase of the pulse. Shifting a phase $\phi$ to the $R_X(\theta)$ pulse in advance yields a $R_\phi(\theta)$ gate. In light of virtual Z gates \cite{McKayPhysRevA17}, this $R_\phi(\theta)$ gate is equivalent to $R_Z(-\phi)R_X(\theta)R_Z(\phi)$. This microwave pulse is the physical realization of $R_{\phi}^{mn}(\theta)$. 

In practice, this method brings two significant sources for errors: coherent and incoherent errors. The former error stems from amplitude miscalibration and has a quadratic impact on algorithmic accuracy. The latter error arises from stochastic noise with linear impact. Those errors can be extracted from the error amplification protocol \cite{SheldonPhysRevA16, gateerrors}. IBM already provides this protocol to fine-tune the amplitude of $\pi$ pulse in subspace \{01\}. To fine-tune the amplitude of $R_X^{12}(\pi)$ pulse, the pulse is repeatedly applied to reveal errors as shown in Fig.~\ref{fig:qutrit} e, where the $x$ axis is the number of pulse $R_X^{12}(\pi)$. The gate sequence is  $[R_X^{12}(\pi)]^n R_X^{01}(\pi)$ so that the qutrit oscillates between state $|1\rangle$ and $|2\rangle$. As seen in Fig.~\ref{fig:qutrit} e, we find an under rotation of 0.008 radians, and a decay rate of 73.125 kHz over the number of pulse gates for qubit 0 of the Jakarta device \cite{SheldonPhysRevA16, gateerrors}. 

\begin{table}[ht]
    \centering
    \begin{tabular}{|c||*{8}{c|}}\hline
    event & figure &  $\varphi_1$ & $\varphi_2$ & $\varphi_3$ & $\varphi_4$ & $\varphi_5$ & $\varphi_6$ & $\varphi_7$ \\\hline\hline
    $\nu_e$ & \ref{fig:vaccum}ab & -1.5312 & -0.4341 & 5.9253, & 6.5312 & -0.4005 & N/A & N/A \\\hline
    $\nu_{\mu}$ & \ref{fig:vaccum}cd, \ref{fig:matter}abc & 1.7018 & -6.2831 & -0.0497 & 3.2981 & -6.4306 & N/A & N/A \\\hline
    $\nu_{\tau}$ & \ref{fig:vaccum}ef & 1.7409 & -0.6074 & -0.6796 & 3.2591 & -0.7130 & N/A & N/A \\\hline
    $\nu_\mu$ & \ref{fig:CP} & -1.9599 & 0.0299 & 0.0299 & 0.0299& 0.0299& -5.8599& 0.0611  \\\hline
    \end{tabular}
    \caption{\textbf{Phase advances between subspaces \{01\} and \{12\}}: Phase advances in radian are tracked using protocols in Eq.~~\eqref{phi6} and ~\eqref{phi8}. These constants are then applied to simulate neutrino oscillations for the vacuum case Eq.~\eqref{RPMNS:vac}, the matter interaction case Eq.~\eqref{R:matter}, and the CP-violation case Eq.~\eqref{R:delta}. The column figure lists the corresponding figure that uses these numbers to calculate their results.}\label{tab:phase}
\end{table}

Thus far, it all works well within one subspace, either \{01\} or \{12\}. However, maneuvering the state in one subspace introduces phase advances in the other. A possible source is the phase accumulation of the idle state when a subspace change occurs. For example, a rotation in subspace \{01\} with duration $t$ imprints on the state $|2\rangle$ a phase proportional to $2\pi(f^{12}-f^{01})t$ \cite{FischerPhysRevResearch22}. Hence, the rotation axis of every gate is modified according to the phase accumulation from previous gates. Since the qutrit starts with state $|0\rangle$, every gate in subspace \{12\} needs a phase correction. Similarly, every gate in subspace \{01\} needs a correction if there is one $R^{12}$ applied earlier. These phase advances depend on unknown parameters, including the detailed design of the qutrit. We manually track and correct them for each qutrit decomposition depending on the number of gates in the circuit. In our analysis, there are two gate decompositions. The vacuum oscillations as in Eq.~\eqref{RPMNS:vac} and the oscillations with matter interaction as in Eq.~\eqref{R:matter} require six qutrit gates, and the CP-violation oscillations as in Eq.~\eqref{R:delta} requires eight qutrit gates. Correspondingly, the gate sequences are modified as
\begin{equation}\label{phi6}
\mathcal{R}^{'}_6 = R^{01}_{\frac{\pi}{2}+\Phi^{01}+\varphi_5} R^{12}_{\frac{3\pi}{2} + \Phi^{12} + \varphi_4}R^{01}_{\frac{\pi}{2} + \Phi^{01} + \varphi_3}R^{01}_{\frac{\pi}{2} + \varphi_2}R^{12}_{\frac{3\pi}{2} + \varphi_1}R^{01}_{\frac{\pi}{2}},
\end{equation}
or
\begin{equation}\label{phi8}
    \mathcal{R}^{'}_8 =R^{01}_{\frac{\pi}{2}+\delta+\Phi^{01}+\varphi_7}R^{12}_{\frac{3\pi}{2}+\Phi^{12}+\varphi_6}R^{01}_{\frac{\pi}{2}+\delta+\Phi^{01}+\varphi_5}R^{01}_{\frac{\pi}{2}+\Phi^{01}+\varphi_4} R^{01}_{\frac{\pi}{2}+\varphi_3}R^{01}_{\frac{\pi}{2}+\delta+\varphi_2}R^{12}_{\frac{3\pi}{2}+\varphi_1}R^{01}_{\frac{\pi}{2}+\delta},
\end{equation}
with $\varphi_i$ being phase advances to each gate due to their previous pulses of the sequence. Since amplitudes are assumed independent of phases, rotation angles are not written explicitly in the above formula. We reconstruct these phases from measurement data. They are estimated to maximize the likelihood that $\mathcal{R}^{'}$ yields the observed probability distribution. These phases are shown in table \ref{tab:phase}, which are fairly stable and do not fluctuate or drift over the course of our analysis. We emphasize that the phase advances for vacuum oscillations and oscillations with matter interaction are identical since their gate decompositions differ only in rotation angles, which are constant numbers. 

To correct for random drifting in IBM's hardware, the first three circuits of any runs are designed for calibration. Their results constitute an inverse matrix for the mitigation protocol \cite{Gambetta17}. Since IBM allows 300 circuits per job, the remaining 297 circuits are used for neutrino simulations. Typically, each job contains 8192 shots and is repeated four times for statistical errors. In all runs, we use real data as much as possible. Their values are $\theta_{12}=33.45^\circ$, $\theta_{23}=42.1^\circ$, $\theta_{13}=8.62^\circ$, $\Delta m_{21}^2=7.42\times10^{-5}\text{eV}^2$, $\Delta m_{31}^2 = 2.510\times10^{-3}\text{eV}^2$. Following famous experiments in the field, we use $E=1$ GeV for neutrino energy when calculating oscillations as a function of distance, and $L=295$ km when calculating oscillations as a function of energy. 

\section{Results and Discussions}

\begin{figure}[ht]
    \begin{center}
        \includegraphics[width=1\textwidth]{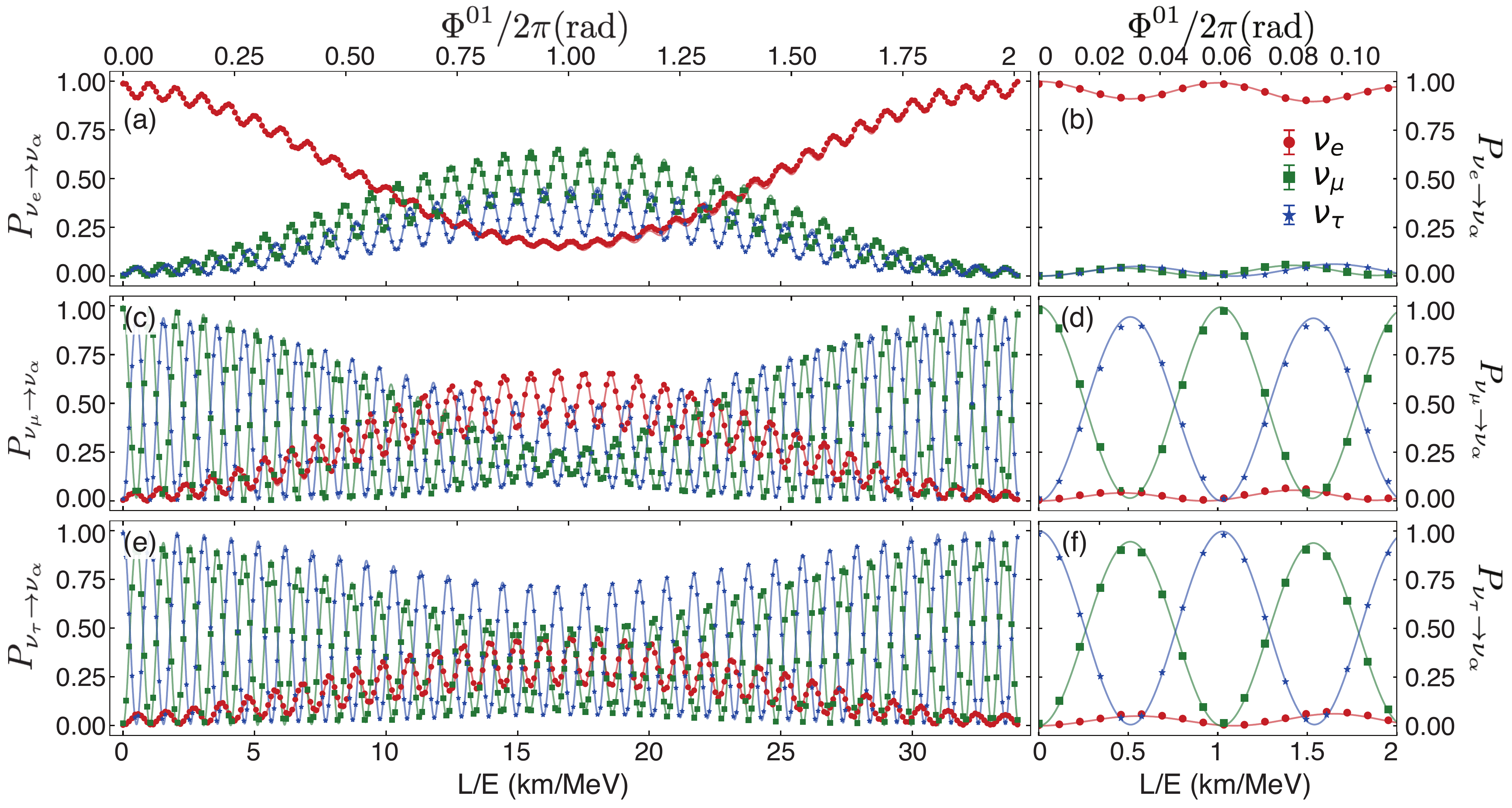}
        \caption{\textbf{Vacuum oscillations:} Survival probabilities when the initial states are (a,b) electron neutrinos, (c,d) muon neutrinos, and (e,f) tauon neutrinos with energy $E=1$ GeV as a function of L/E in an entire period of $\Phi^{01}$. The solid lines are theoretical calculations from classical computers using Eq.~\eqref{eq:classical_prob}, while the dots are quantum simulation results on a real quantum computer following Eq.~\eqref{RPMNS:vac}. Panels (b, d, f) zoom in to a small regime of the corresponding graphs on their left. Each dot is averaged from 4 runs with 8192 shots each. Error bars are smaller than their symbols and are barely visible.}
        \label{fig:vaccum}
    \end{center}
\end{figure}

With calibrated qutrit pulses, we implement the PMNS actions by gating a qutrit and thus simulate neutrino oscillations on IBM quantum hardware. Specifically, neutrino oscillations in vacuum are simulated as a chain of 6 qutrit gates on Jakarta qubit 0 per Eq.~\eqref{RPMNS:vac}. In Fig.~\ref{fig:vaccum}, the probability of detecting different neutrino flavors is presented. From top to bottom panels, the initial states of neutrino are chosen as electron neutrinos $|0\rangle$, muon neutrinos $|1\rangle$, and tauon neutrinos $|2\rangle$ on Fig.~\ref{fig:vaccum} a and b, Fig.~\ref{fig:vaccum} c and d, and Fig.~\ref{fig:vaccum} e and f, respectively. The left column shows a full period, and the right column shows a zoom-in of the same data. The probabilities for final state  $\nu_e$, $\nu_{\mu}$, and $\nu_{\tau}$ are in red, green, and blue, respectively. In all graphs, simulated data are dots with error bars barely visible, and analytical calculations using the PMNS matrix are solid lines. The bottom $x$ axes show the standard scale in neutrino studies in length per energy $L/E$ with $E = 1$ GeV and $L=t$ in the relativistic scale. The top $x$ axes show the rotation axes $\Phi^{01}/2\pi$ used for the qutrit according to Eq.~\eqref{Phi01}. All curves are executed four times independently with 8192 shots each. Dots from Fig.~\ref{fig:vaccum}a, c, and e are averaged from these runs with relative errors mostly in the range from 1 to 10\%. 

\begin{figure}[th]
    \begin{center}
    \includegraphics[width=0.6\textwidth]{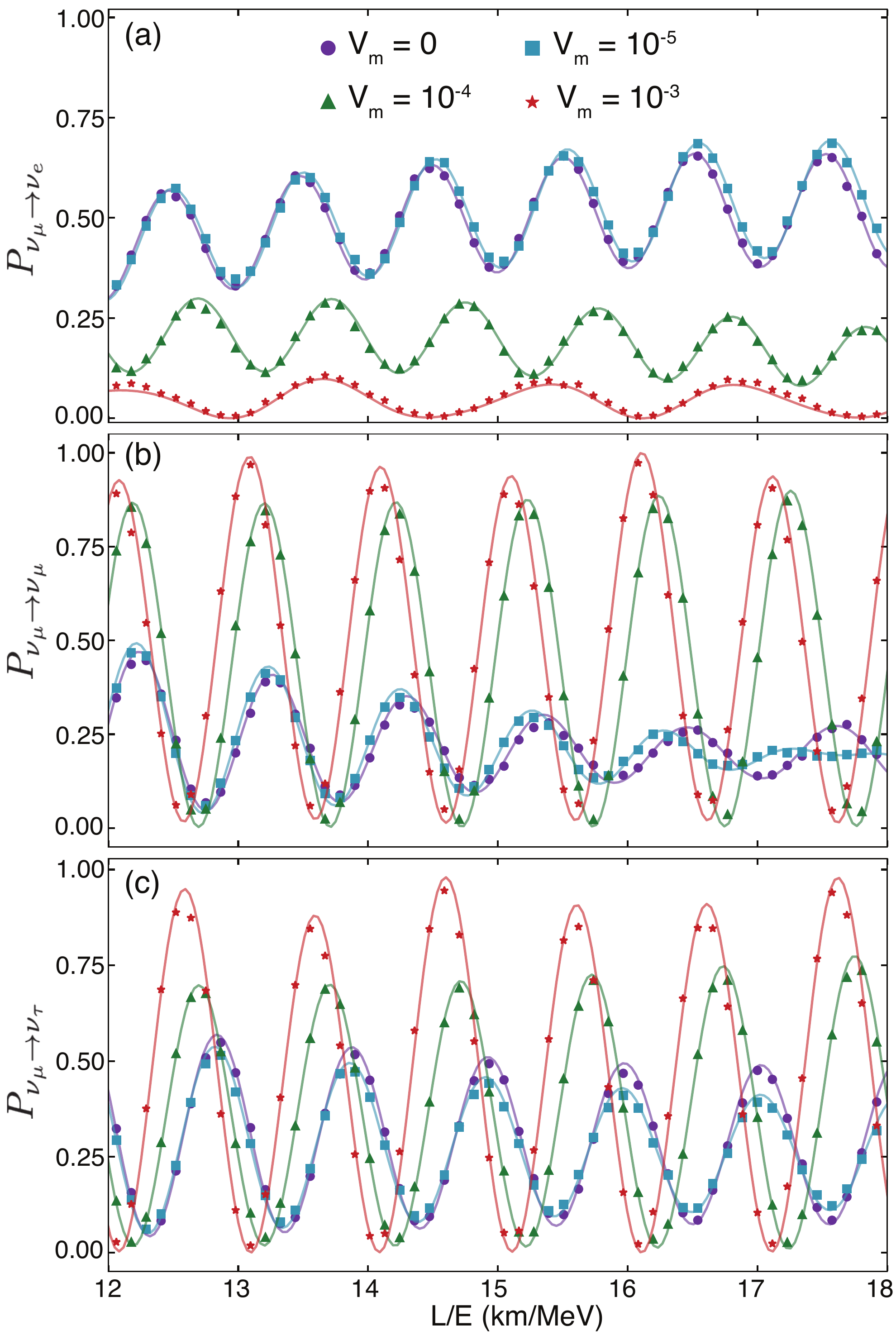}
    \caption{\textbf{Oscillations with matter interactions:} Oscillations of a muon neutrino with E = 1 GeV traveling a distance L in matter with potential in unit $\text{eV}^2$: $V_m = 0$ in purple, $10^{-5}$ in cyan, $10^{-4}$ in green, and $10^{-3}$ in red. Comparison of matter effect for four potentials is shown in three channels where an initial muon neutrino oscillates to electron neutrino (a), muon neutrino (b), and tauon neutrino (c). Solid lines are PMNS analytical calculations. Dots indicate data from the IBM quantum computer.}\label{fig:matter}
    \end{center}
\end{figure}

In a similar manner, neutrino oscillations when interacting with matter are simulated, as shown in Fig.~\ref{fig:matter}. Assuming only interactions with electrons via the potential Eq.~\eqref{Matrix:Vc}, the Hamiltonian contains a correction term $V_m$ as a perturbation. Following the diagonalization as in Eq.~\eqref{PMNS:matter}, the PMNS matrix is decomposed similarly to the vacuum case into a sequence of 3 qutrit rotation gates. This decomposition Eq.~\eqref{R:matter} has the same form as the decomposition for vacuum oscillations Eq.~\eqref{RPMNS:vac}. All calculations here, therefore, resemble the case of vacuum oscillations, with some modifications to all constants. Without CP violation $\delta = 0$, four different values for $V_m = 0, 10^{-5}, 10^{-4}$, and $10^{-3}$ eV$^2$ are chosen to simulate the oscillations with muon neutrino $|1\rangle$ as the initial state. On Fig.~\ref{fig:matter}, the oscillation probabilities in three channels $\nu_\mu \rightarrow \nu_e$, $\nu_\mu \rightarrow \nu_{\mu}$, and $\nu_\mu \rightarrow \nu_{\tau}$ are presented. In each channel, the probability of detecting a flavor is shown with four values of matter potentials, $V_m = 0$ eV$^2$ in purple, $10^{-5}$ eV$^2$ in cyan, $10^{-4}$ eV$^2$ in green, and $10^{-3}$ eV$^2$ in red. Each curve is averaged from 8192 shots. Following Eq.~\eqref{phi6} and table \ref{tab:phase}, the same phase corrections between qutrit gates in Eq.~\eqref{R:matter} as in vacuum oscillations are used. In all cases, data from real hardware matches well with analytical calculations. The relative errors are mostly around 1\% to 10\%, similar to the case of vacuum oscillations.

\begin{figure}[ht]
    \begin{center}
    \includegraphics[width=0.7\textwidth]{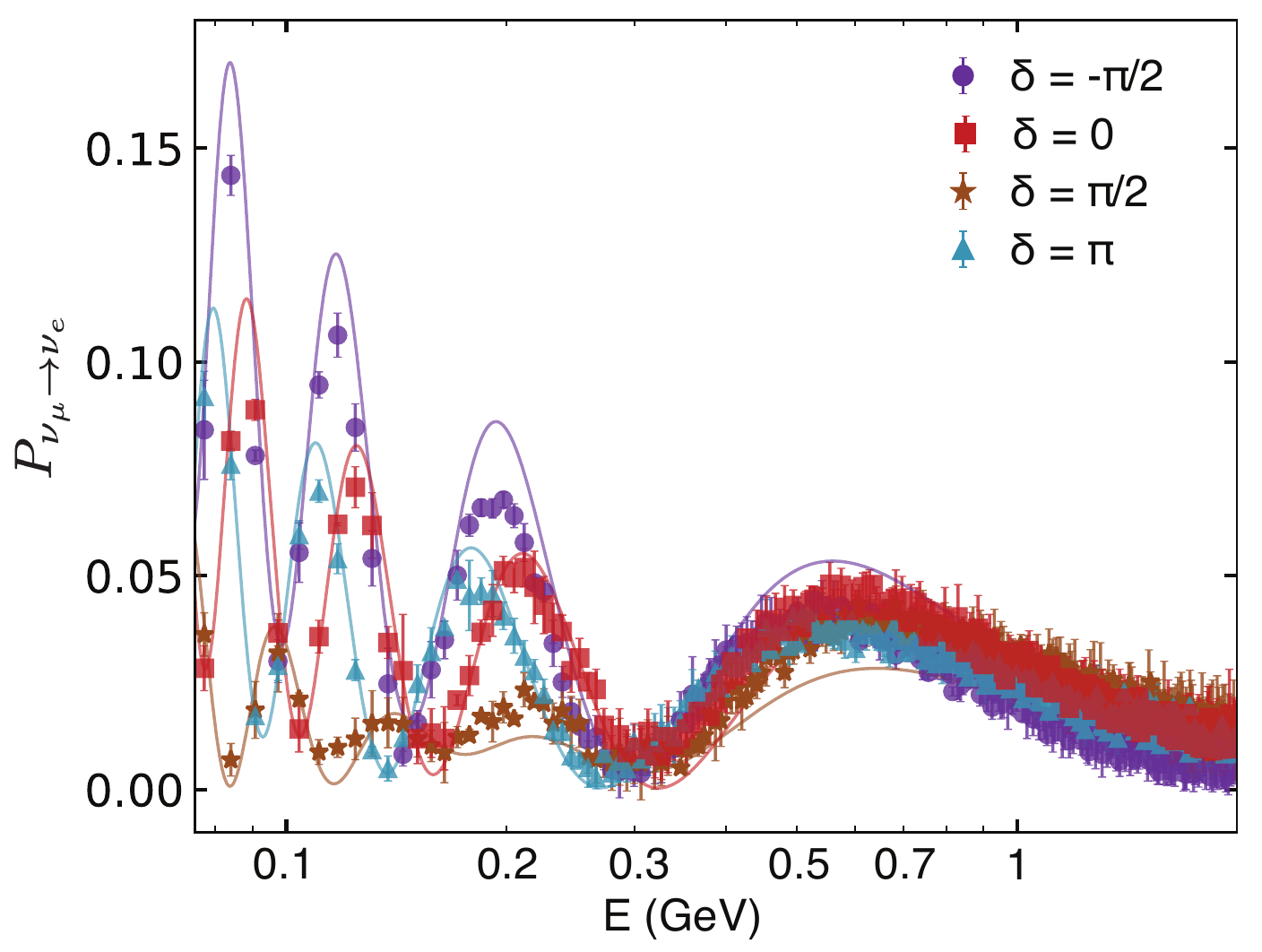}
    \caption{\textbf{Oscillations with CP-violation:} 
    Calculations of the appearance probability of electron flavor from an initial muon neutrino as functions of energy E in four notable cases of CP violating phases, $\delta = \pi/2,\pi, 0, -\pi/2$. The baseline is $L=$ 295 km, representing T2K configuration; matter effects are not considered with $V_m=0$. The solid lines are analytical calculations using Eq. \eqref{eq:classical_prob}, while the dots are quantum simulation outputs from the IBM quantum computer}\label{fig:CP}
    \end{center}
\end{figure}

To demonstrate the power of our approach to the PMNS theory, neutrino oscillations are further simulated in the presence of CP-violation $\delta\neq 0$ to the $\sin(\theta_{13})$ term in Eq.~\eqref{eq:UPMNS3}. Unlike the case of vacuum oscillations where $\delta = 0$ in the decomposition Eq.~\eqref{RPMNS1}, the term $\theta_{12}$ does not share a common rotation axis with the term $\alpha_3$. The PMNS matrix is decomposed into four rotations of different axes, as shown in Eq.~\eqref{RPMNS:delta}. There are seven phase advances $\varphi_i$ corresponding to eight axes of gate sequence as in protocol Eq. ~\eqref{phi8}, which is given in table \ref{tab:phase}. In Fig.~\ref{fig:CP}, we calculate the probability for the oscillation channel from $\nu_{\mu}$ to  $\nu_{e}$ in four cases: maximum neutrino enhancement with $\delta = -\pi/2$, maximal anti-neutrino enhancement $\delta = \pi/2$, $\delta = 0$, and $\delta = \pi$. The oscillation is now plotted as a function of energy at distance $L = 295$ km, corresponding to the configuration of the T2K experiment \cite{T2K2011}. Each curve is averaged from 4 datasets with 4096 shots. Data for the other two cases $\nu_\mu \rightarrow \nu_{\mu}$ and $\nu_\mu \rightarrow \nu_{\tau}$ can be found on our GitHub repository \cite{Github}. The probability in the $\nu_\mu \rightarrow \nu_e$ channel is smaller than that of the other two cases, thus leading to larger relative errors. Still, the relative errors for all data are mostly in the range of 1 to 10\%. Extending these simulations for other flavors, different distances, or values for $\delta$ is straightforward. 

\begin{table}[ht]
    \centering
    \begin{tabular}{|l||*{3}{c|}|*{4}{c|}|*{4}{c|}}\hline
        & \multicolumn{3}{|c||}{vacuum (Fig.~\ref{fig:vaccum})} & \multicolumn{4}{|c||}{$V_m$ (eV$^2$) (Fig.~\ref{fig:matter})} & \multicolumn{4}{|c|}{$\delta$ (Fig.~\ref{fig:CP}) }\\\cline{2-12}
        & $\nu_e(0)$ & $\nu_\mu(0)$ & $\nu_\tau(0)$ & 0 & $10^{-5}$ & $10^{-4}$ & $10^{-3}$ & $-\frac{\pi}{2}$ & 0 & $\frac{\pi}{2}$ & $\pi$\\\hline\hline
        $\nu_e$ & 0.9983 & 0.9975 & 0.9965 & 0.998 & 0.999 & 0.983 & 0.927 &  0.967 & 0.980 & 0.924 & 0.998\\\hline
        $\nu_\mu$ & 0.9976 & 0.9955 & 0.9985 & 0.993 & 0.996 & 0.996 & 0.987  & 0.955 & 0.964 & 0.963 & 0.971 \\\hline
        $\nu_\tau$ & 0.9947 & 0.9955 & 0.9994  & 0.992 & 0.994 & 0.996 & 0.987 & 0.947 & 0.959 & 0.966 & 0.974\\\hline
        \end{tabular}
        \caption{\textbf{$R^2$ scores:} calculated using Eq.~\eqref{eq:R2} to compare theoretical calculation from PMNS theory and simulation data from quantum computers for all data presented in this work: vacuum oscillations shown in Fig.~\ref{fig:vaccum}, oscillations when interacting with matter shown in Fig.~\ref{fig:matter}, and oscillations in the present of CP-violation shown in Fig.~\ref{fig:CP}.}
\label{tab:R2}
\end{table}

To compare our quantum simulation approach with traditional analytical calculations, we calculate the $R^2$ score and the relative error following their definitions
\begin{equation}\label{eq:R2}
    R^2 = 1-\frac{\sum_{i=0}^{N-1}(y_i-y_{0})^2}{\sum_{i=0}^{N-1}(y_i-\bar{y})^2},
\end{equation}
and
\begin{equation}
    \Delta y = \frac{|y_i-y_0|}{|y_0|}.
\end{equation}
Here, $y_{0}$ are the theoretical probabilities, $y_i$ are the experimental data, and $\bar{y}$ are their means. As seen in table \ref{tab:R2}, $R^2$ are mostly bigger than 99\%, indicating a great fit between the analytical approach and the quantum simulation on real qutrit. The value for relative errors $\Delta y$ are shown in all figures. With a typical range from 1 to 10\%, these error bars are barely visible.

Apparently, encoding three neutrino flavors into qutrit eigenstates has certain advantages over the qubit approach \cite{MolewskiPRD22}. Instead of using two entangled qubits, utilizing the qutrit does not involve a redundant basis. The PMNS matrix is expressed only with single qubit gates in the two subspaces without any complicated entanglement gates. In previous simulations using qubits \cite{MolewskiPRD22}, the PMNS matrix Eq.~\eqref{eq:UPMNS3} is decomposed as a product of qubit rotations $R_{\rm{PMNS}} = R_{23}(\theta_{23}, 0) R_{13}(\theta_{13},\delta)R_{12}(\theta_{12}, 0).$ On the two-qubit Hilbert space, $R_{12}$ rotation is constructed from the native Controlled-U3 gate. The other rotations $R_{23}$ and $R_{13}$ are constructed from this base by adding permutation matrices such as SWAP gates. In total, the qubit decomposition comprises 3 Controlled-U3 gates, 2 CNOT gates, and 4 SWAP gates. Even after simplification, the PMNS decomposition on a two-qubit system requires 3 Controlled U3 and 2 CNOT gates, which is quite a burden for current quantum hardware. Compiling this circuit to pulse schedules, the execution time to implement qubit-based PMNS on IBM devices is $12224~dt$, in comparison to our execution time using qutrit is $640~dt$ per Eq.~\eqref{RPMNS:vac}. The qutrit-based approach is 19.6 times faster compared to the qubit-based calculation. 

The main challenge in our work relates to engineering the qutrit. Different from the popularity of qubits, qutrits get little attention, and there is much work to be done. To reduce SPAM errors, we have to scan a wide range of amplitude and duration, which constructs a silhouette score heat map that navigates us to the best pulse parameters. To prevent drifting, every job has its mitigation matrix. Errors of single pulse gate in \{01\} and \{12\} subspace is maintained in the range of $10^{-3}$ to $10^{-4}$ and $10^{-3}$ to $10^{-2}$, respectively.  However, the most challenging issue is the unknown nature of the correlation between the two subspaces \{01\} and \{12\} of the qutrit. We obtain constant phase advances between subspaces by testing identical gate sequences to the oscillation quantum circuits. These numbers remain unchanged, as shown in table \ref{tab:phase}. A more systematic investigation is currently going on, with results lying outside the scope of this work. In the near future, optimal and robust control techniques \cite{BaumPRXQuantum21, WerninghausNPJ2021, Carvalho_PRA2021, Wu_PRL2020} may improve qutrit gate fidelity, especially for $\delta \neq 0$ circuits.

Our simulations demonstrate the complex interplay among neutrino parameters in terms of well-controlled pulse parameters on real quantum hardware. With a diagonal Hamiltonian, neutrino evolutions can be decomposed into sequences of rotations in the qutrit space. The oscillations can be reproduced with precision both without and with matter interactions. Moreover, CP-violation physics $\delta\neq 0$ can be incorporated into the qutrit circuit as a modification in microwave phases. Different interpretations of CP-violation physics can be achieved by extending Eq.~\eqref{eq:UPMNS3} to include different parameters. For example, the case when $\delta$ associates with other terms, say $\sin\theta_{12}$ or $\sin\theta_{23}$. It is straightforward to extend our results to an arbitrary dimensional system of many neutrinos with a multi-qutrit system \cite{Goss2022}. The genuine qutrit high-dimensional entanglement may benefit classically intractable problems such as out-of-equilibrium dynamics of collective neutrinos \cite{amitrano2023trapped}.

In summary, our replication of the PMNS theory on a generic qutrit demonstrates that quantum computers are a valuable tool for studying neutrino physics. We engineer a qutrit with high-quality control and measurement from a transmon device available on the cloud. Reliable single qutrit operations are achieved using low-level microwave controls. With error mitigation and careful calibrations, the processes are stable and accurate for every run. Our circuit is short and precise with efficient encoding, resulting in improved leakage error and lesser system drifting. Three scenarios for the neutrino oscillations have been simulated, including vacuum oscillations, interaction-with-matter oscillations, and CP-violation oscillations. Our simulations match well with the PMNS theory and state-of-the-art experiments in all cases.

\section*{Acknowledgement}
We acknowledge the support of Qiskit Slack team, especially Thomas Alexander. We appreciate Isha Mehta for the contribution in the early stage. We would like to thank Van-Nam Do for the fruitful discussions.
\bibliographystyle{apsrev4-2}
\bibliography{ref}
\end{document}